\documentclass[aps,prl,twocolumn,showpacs,superscriptaddress]{revtex4}

\usepackage{graphicx}
\usepackage{float}

\begin{document}
\title{One-dimensional conduction in Charge-Density Wave nanowires}

\author{E. Slot}
\author{M. A. Holst}
\author{H. S. J. van der Zant}
\affiliation{Kavli Institute of NanoScience Delft, Delft University of Technology, Lorentzweg 1, 2628 CJ Delft, The Netherlands}
\author{S. V. Zaitsev-Zotov}
\affiliation{Institute of Radioengineering and Electronics, Russian Academy of Sciences, Mokhovaya 11, 125009 Moscow, Russia}
\date{\today}

\begin{abstract}
We report a systematic study of the transport properties of coupled one-dimensional metallic chains as a function of the number of parallel chains. When the number of parallel chains is less than 2000, the transport properties show power-law behavior on temperature and voltage, characteristic for one-dimensional systems.
\end{abstract}



\maketitle
Electron-electron interactions in one-dimensional (1D) metals exhibit dramatically different behavior from three-dimensional metals, in which electrons form a Fermi liquid. Depending on the details of the electron-electron interaction, several phases are possible, such as a Luttinger Liquid (LL) or a 1D Wigner crystal (WC). The tunneling density-of-states (TDOS) of these 1D phases exhibit power-law behavior on the larger one of either $eV$ or $k_BT$ \cite{voit,lee,glazman}, with $V$ the voltage and $T$ the temperature.

This power-law behavior has been observed in various 1D systems, such as ballistic single-wall \cite{bockrath} and diffusive multi-wall carbon nanotubes \cite{Bachtold}, degenerately doped semiconductor nanowires \cite{Sergei}, and fractional quantum Hall edge states \cite{Chang}. Each of these systems revealed new behavior, but also raised questions. For example in multiwall carbon nanotubes (MWCN), the different nanotube shells interact with each other and the question arose how this interaction affects the 1D properties. It still under debate whether the observed power-law behavior in MWCNs should be described by LL or by Environmental Coulomb Blockade Theory (ECBT) \cite{Bachtold}.

Theoretically, it has been shown that the LL state survives for a few 1D chains coupled by Coulomb interactions \cite{mukhopadhyayvishwanath}. However, when \textit{interchain} hopping is taken into account, the LL state is destroyed for low temperatures \cite{biermann}. It is less clear what the situation is when many $(>10)$ chains are coupled together in the presence of disorder. A single impurity in a 1D chain forms a tunnel barrier \cite{kane}. More impurities in coupled chains may have different consequences. Recent theoretical investigations show that a LL state can also be stabilized in the presence of impurities for system with more than two coupled chains \cite{artemenko}. Furthermore, formation of a WC is expected in the limit of strong Coulomb interactions, or low electron density \cite{lee,giamarchi}. All these considerations lead to a large parameter space, and at present a full theory of 1D transport in disordered, multichannel systems is still lacking.

In this paper, we show that charge-density wave (CDW) nanowires exhibit the characteristic behavior for 1D transport. We use the model CDW compound NbSe$_3$, which consist of metallic chains weakly coupled by Van der Waals forces. Nanowires consisting of more than thousands of chains are metallic down to the lowest temperatures, in accordance with the bulk behavior of NbSe$_3$ whiskers. In this limit, CDWs can be viewed as the classical analogue of a LL-state. However, we find that at low temperatures our nanowires with less than 2000 chains become insulating at low temperatures. Power-law dependencies on both voltage and temperature are observed, characteristic for 1D transport.

The monoclinic unit cell of NbSe$_3$ contains 6 metallic chains along the $b$ direction. Its lattice parameters are $a = 10.0$~\AA, $b = 3.5$~\AA, $c = 15.6$~\AA, and $\theta = 109.5^o$ \cite{Smaalen}. NbSe$_3$ is a metallic CDW material with a partially gapped Fermi surface. Two Peierls transitions occur at $T_{P1} = 145$~K and $T_{P2} = 59$~K. Below the second Peierls transition nearly all of the electrons are condensed into both CDWs. The remaining uncondensed electrons have a very low electron density of $n=1.1\times 10^{18}$~cm$^{-3}$ \cite{Ong}, indicating that electron-electron interactions can become important.

\begin{figure}
\includegraphics*[width=7.5cm, bb = 1 1 719 253]{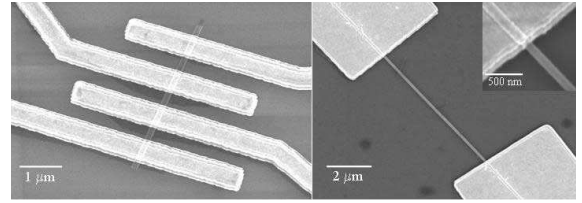}
\caption{\label{sem}Scanning Electron Microscope images of NbSe$_3$ nanowires. Left: A four-probe device. Right: A two-probe device. Right inset: A zoom-in on the contact of the two-probe device.}
\end{figure}

\begin{figure}
\includegraphics*[angle=0,width=7cm]{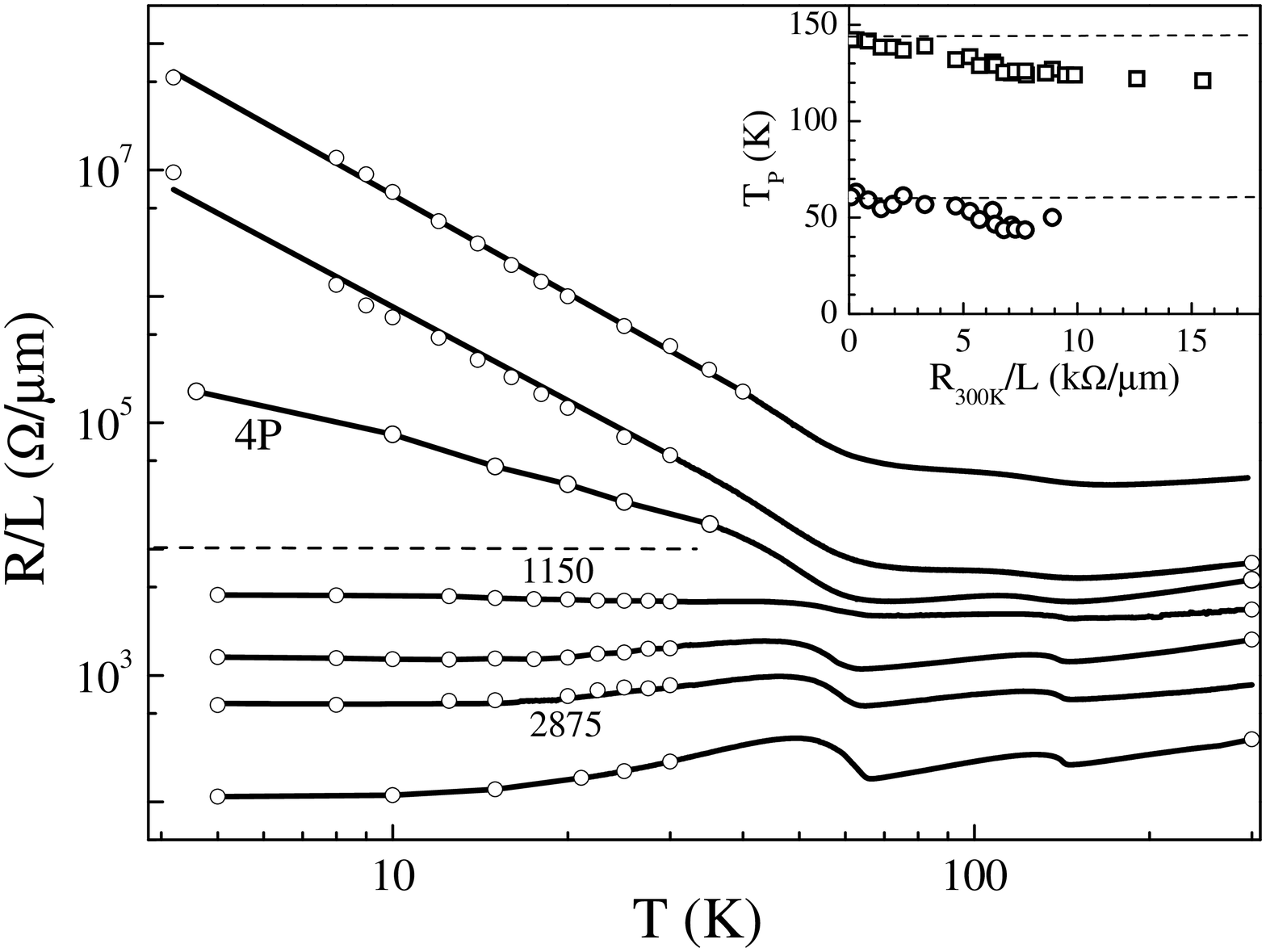}
\caption{\label{RLvsT}Resistance per unit length $R/L$ as a function of $T$ on a log-log scale. Small nanowires (high $R_{300K}/L$ value) show metallic behavior at room temperature, but show non-metallic behavior at low temperature. All curves are two-probe (2P) measurements except for the one indicated by 4P. The open circles are data points taken manually from $IV$ curves. The number of \textit{participating} chains in CDW conduction is deduced from Shapiro-step measurements at $T=120$~K and indicated in the figure for two samples. Inset: Peierls transition temperature as a function of the cross section.}
\end{figure}

NbSe$_3$ nanowires were made from bulk NbSe$_3$ crystals by ultra-sonically cleaving the crystals in a pyridine solution. After several hours of cleaving, a suspension of NbSe$_3$ nanowires with widths ranging from 30~nm to 300~nm and lengths from 2 to 20~$\mu $m, emerges. A drop of the suspension is deposited onto a degenerately doped Si substrate with predefined markers. The Si can serve as a backgate, but is connected to ground in the measurements shown here. Nanowires are selected and located with an optical microscope with respect to the predefined markers. Subsequently a contact pattern is defined with e-beam lithography. Gold and a Ti sticking layer are deposited within minutes after a 3 second dip in ammonium buffered hydrofluoric acid to optimize contact resistances. Both two-probe and four-probe samples were made, see Fig.~\ref{sem}. In total, 21 nanowires have been measured. One wire (4P sample in Fig. \ref{RLvsT}) has also been measured after prolonged exposure to air. We found that the room-temperature resistance of this nanowire monotonically increased with exposure time (a factor 2 of resistance increase after 1100 hours). Most likely, the surface slowly oxidizes thereby reducing the number of chains that participate in conduction.

Figure \ref{RLvsT} shows the zero-bias resistance $R$ per unit length $L$ as a function of $T$ for several nanowires plotted on a log-log scale. All samples, both large and small, show metallic behavior at room temperature. The samples with the largest cross sectional areas are at the bottom of the graph and show the two Peierls transitions clearly. These samples remain metallic down to $T=4.2$~K, similar to bulk crystals. The samples with the smallest cross sectional areas are at the top of the graph. When the room-temperature unit-length resistance $R_{300K}/L$ is larger than 10~k$\Omega /\mu$m, the Peierls transitions become increasingly less visible while the transition temperatures show a small monotonic decrease of up to 20\% for the smallest sample, see inset of Fig.~\ref{RLvsT}.

The most remarkable behavior is the change from metallic to non-metallic behavior at low temperatures when the number of parallel chains decreases. This transition occurs around $R_{300K}/L = \rho / A = 4$~k$\Omega /\mu$m corresponding to a cross section $A=500$~nm$^2$ and a total number of 2000 chains (bulk NbSe$_3$ resistivity $\rho=2$~$\Omega \mu$m).

To characterize the sample quality, we have performed Shapiro-steps measurements at $T=120$~K on two samples that show the Peierls transitions. We were able to mode-lock the CDW completely to an external radio-frequency signal on both samples. From the distance between mode-locking steps we deduced the number of chains participating in CDW conduction, indicated by the numbers in Fig.~\ref{RLvsT}. Two out of six chains participate in CDW conduction and the cross sectional area deduced from the number of chains agrees to within 10\% of the value obtained from room-temperature resistance. This observation indicates that the wires are homogeneous and that the contact resistance at $T=120$~K is small compared to the nanowire resistance.

Of the 7 curves presented in Fig.~\ref{RLvsT}, one has been obtained from a four-probe measurement, indicated by 4P. For this sample, the 4P resistance is equal to the two-probe resistance to within 10\% over the entire temperature range 4.2~K to 300~K. The interface of the contact and the nanowire is transparent and therefore, the non-metallic behavior at low temperature seen for this sample is not an effect of the interface.

\begin{figure}[b]
\includegraphics*[angle=0,width=7cm]{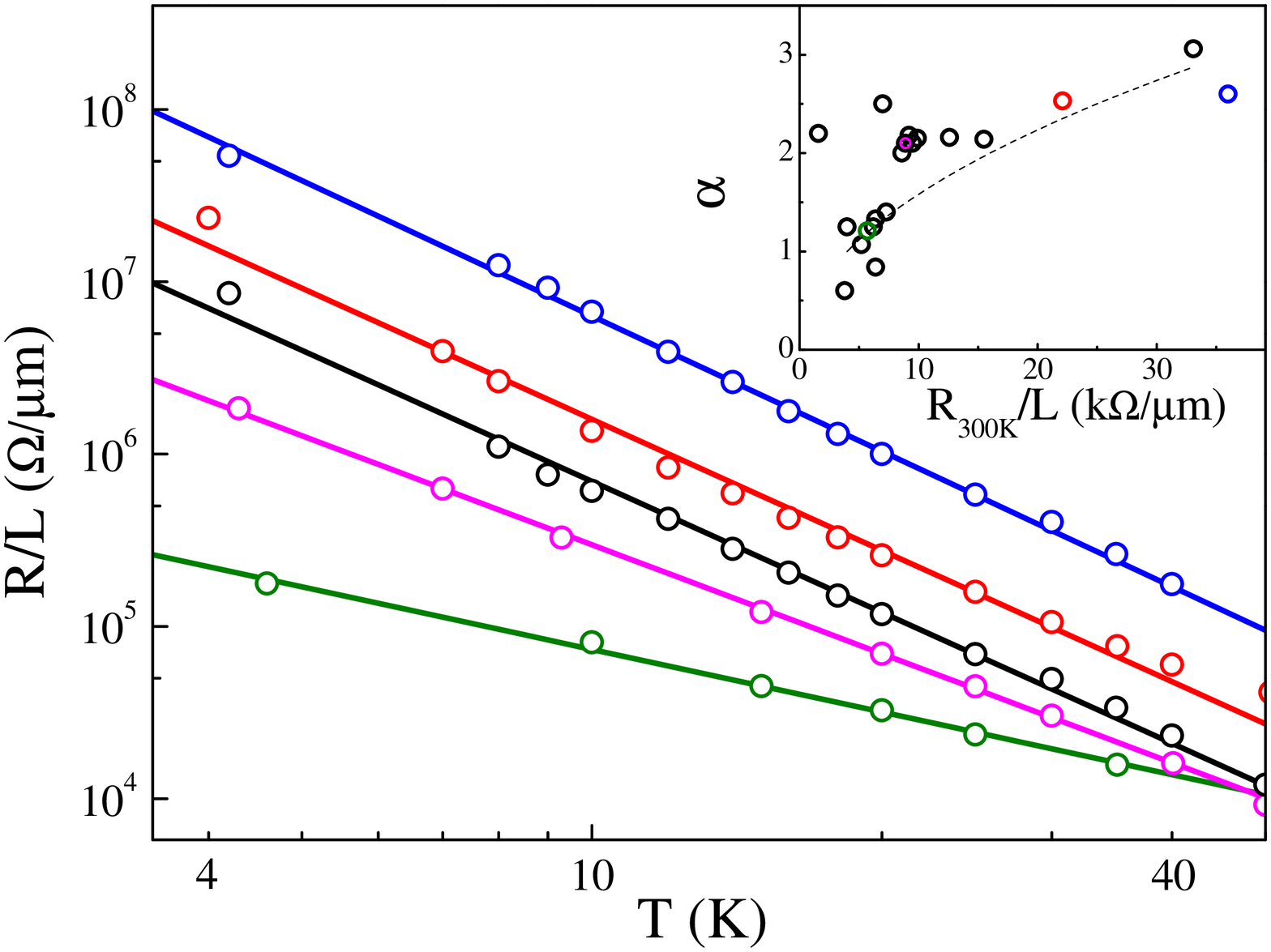}
\caption{\label{RT} $R/L$ as a function of $T$ on a log-log scale. The lines are linear fits to the data from which the power-law exponent $\alpha$ is deduced. The inset shows the exponents for all samples measured as a function of the room-temperature resistance per unit length.}
\end{figure}

At low temperature for all samples with $R_{300K}/L > 5$~k$\Omega/\mu$m, power-law behavior $R\propto T^{-\alpha}$ is observed, as illustrated in Fig.~\ref{RLvsT} and in Fig.~\ref{RT} for five other samples. In the inset of Fig.~\ref{RT} we have plotted all the $\alpha$-values as a function of $R_{300K}/L \propto 1/A$. The exponents seem to increase as the number of chains becomes smaller, although on the other hand, the exponents can also be divided into two groups around $\alpha=1$ and $\alpha=2$. The $R(T)$ curves have also been fitted to the thermal activation and Mott hopping law $R \propto \text{exp}(T_0/T)^\delta$, where $\delta=1$ for thermal activation and $\delta$ depends on dimensionality for Mott hopping \cite{Fogler}. Both models do not fit the data satisfactorily.

\begin{figure}
\includegraphics*[angle=0,width=7cm]{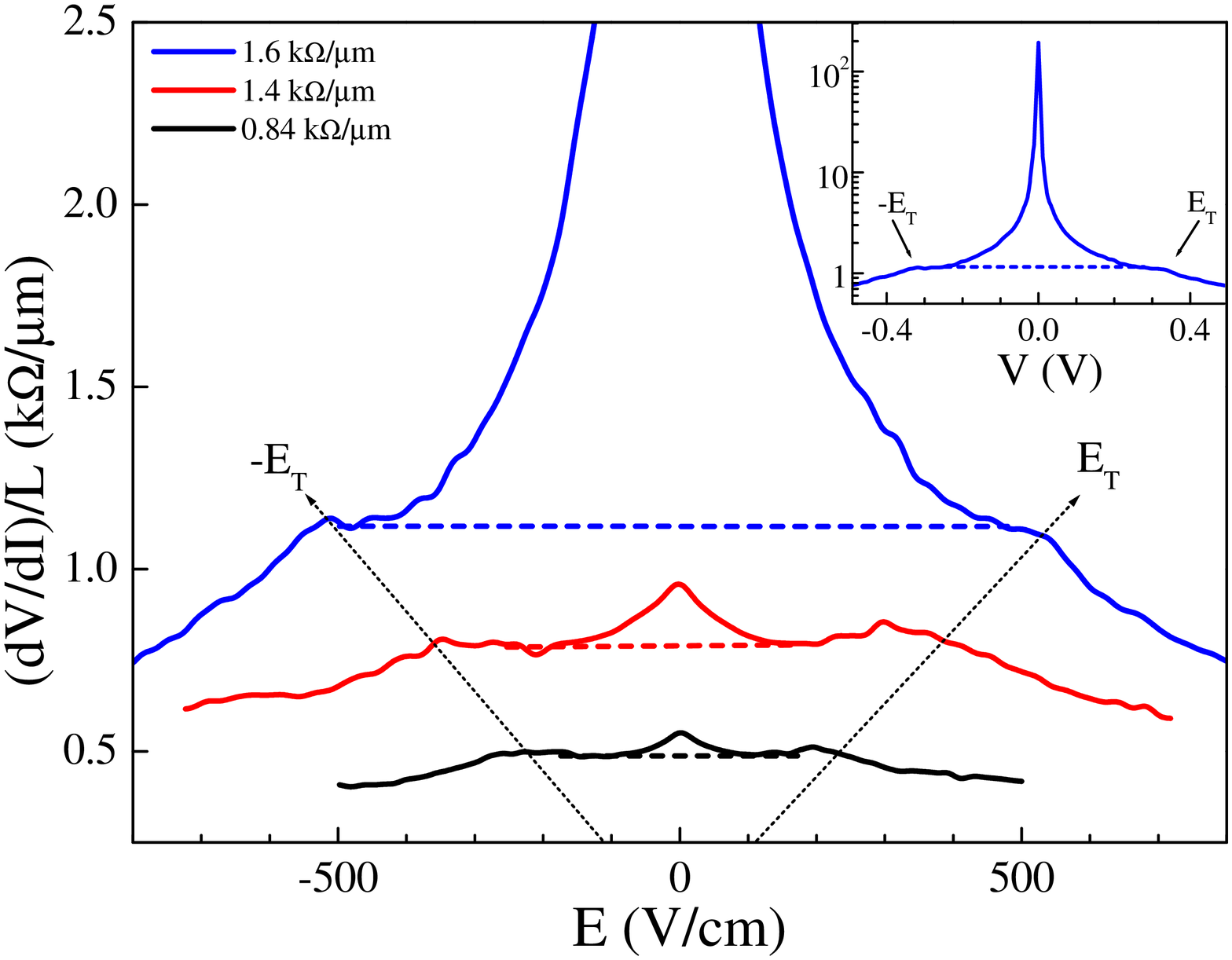}
\caption{\label{dVdI} Differential resistance per unit length $(dV/dI)/L$ as a function of the voltage per unit length $E$ of three nanowires at $T=4.2$~K. The threshold field for sliding $E_T$ increases when the cross sectional areas gets smaller, indicated by the dashed arrows. Below $E_T$ where the CDW is pinned, a peak in $dV/dI$ is observed. The dashed lines indicates the shape of the $dV/dI$ expected for bulk samples.
The room-temperature unit length resistance is displayed for each curve. The inset shows the $(dV/dI)/L$ of the top curve as a function of the voltage $V$; the maximum resistance is 190 k$\Omega/\mu$m and $L=6.2~\mu$m.}
\end{figure}

Figure \ref{dVdI} shows the differential resistance per unit length $(dV/dI)/L$ as a function of the voltage per unit length $E$ at $T=4.2$~K for three samples. The nanowire with the smallest cross section is at the bottom of the graph. When the cross section gets smaller, the threshold field for sliding $E_T$ increases as expected \cite{erwin}. A peak in the differential resistance develops for $E < E_T$, which grows as the cross section gets smaller. The top $(dV/dI)/L$ curve follows a power-law $I \propto V^\beta$ at high bias, with $\beta=1.7$ and on temperature with $\alpha=2.2$. Therefore, power-law behavior is observed simultaneously with CDW features. In this respect our results are completely different from earlier attempts to observe 1D transport in CDW conductors \cite{sergei2}, where complete disappearance of both transitions was reported.

\begin{figure}[b]
\includegraphics*[angle=0,width=7.5cm]{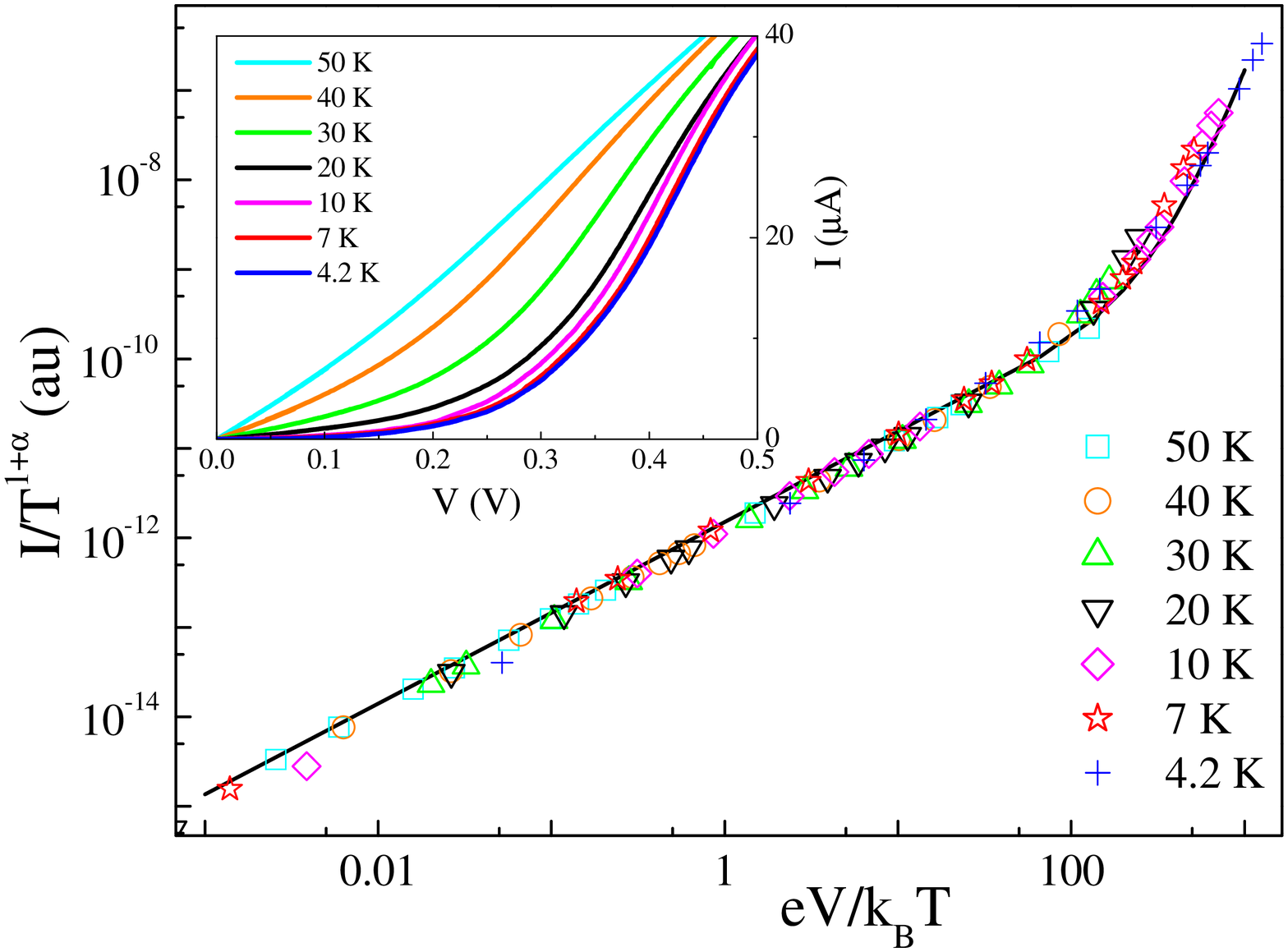}
\caption{\label{IV} Scaled $IV$ curves on a log-log scale. The $IV$s collapse onto a universal curve from $T=4.2$~K to $T=50$~K for $\alpha=2.15$. The inset shows the unscaled $IV$s ($L=1.4~\mu$m).}
\end{figure}

Figure~\ref{IV} shows the current-voltage ($IV$) curves for one sample at several temperatures. At $T=50$~K, the $IV$ is almost linear, but at $T=4.2$~K the $IV$ is highly non-linear. All $IV$ curves collapse unto a single master curve, by plotting $I/T^{1+\alpha}$ versus $eV/k_BT$, where $\alpha=2.15$ is the power-law exponent from the $R(T)$ curve. The scaled $IV$ curves are non-linear above $eV \approx 80~k_BT$, and for $eV > 80~k_BT$, the $IV$ follows a power-law with $\beta = 4.2$ for this sample. It is important to note that sliding of the CDW occurs at higher voltages \cite{erwin} as is evident from the measurement at $T=50$~K. At this temperature, the threshold field for sliding is the lowest and we still observe linear behavior in the inset of Fig.~\ref{IV}.

The scaled master curve can be fitted to a general equation to describe $IV$s for bosonic excitations in 1D \cite{Balentsvenkataramanthesis}:
\begin{equation}
\label{bosonequation}
\frac{I}{T^{1+\alpha}} = C \sinh \left( \gamma \frac{eV}{k_BT} \right)\left|\Gamma\left(1+\frac{\beta}{2}+i \gamma \frac{eV}{\pi k_BT} \right)\right|^2,
\end{equation}
where $\Gamma $ is the complex gamma function, $C$ is a proportionality constant and $\gamma $ determines the position of the `knee' in the $IV$ curve. The parameters $\alpha$ and $\beta$ are the two experimentally determined exponents for the temperature and voltage dependence respectively. The fit to the data is depicted by the solid line in Fig.~\ref{IV}, with $C = 2.3 \times 10^{-11}$ and $\gamma = 77^{-1}$. $IV$s have also been measured for another sample with $\alpha=2.16$ and $\beta=4.7$ and again a similar fit could be made. The fit parameters for this sample are $C = 1.1 \times 10^{-11}$ and $\gamma = 100^{-1}$.

The power-law behavior in NbSe$_3$ nanowires results from a reduction of cross section below about 2000 parallel chains. The data clearly show that CDW sliding does not play a role. Power-law dependence is observed below $T=50$~K suggesting that the low electron density below $T_{P2}$ is important. The measurements show that these carriers dominate transport and govern the metallic to nonmetallic transition. The CDW state still exists, because power-law behavior is observed simultaneously with the threshold field for CDW sliding (Fig.~\ref{dVdI}). Furthermore, $T_{P1,2}$ decreases only slightly with the number of chains (inset of Fig.~\ref{RLvsT}). This is in agreement with the expectation that already a small number of chains ($\approx 10$) stabilizes the CDW state \cite{artemenkoprivate}. In the remainder of this paper, we discuss the 1D transport mechanism and compare our results with different models.

When comparing NbSe$_3$ nanowires to other systems that show power-law behavior, we have to take into account that NbSe$_3$ is a diffusive conductor. Therefore comparison to a single channel LL without disorder, such as SWCNs, is inappropriate. We also have to take interaction between chains into account. A suitable system to compare NbSe$_3$ nanowires to is MWCNs, which are diffusive conductors with interaction between the nanotube shells. MWCN have been modelled using ECBT \cite{Bachtold,Egger}. The $IV$ follows a power-law on energy, with an exponent $\beta=\frac{2Z}{R_Q}+1$ that depends on the impedance $Z$ of the environment \cite{glazman}. Here, $R_Q$ is the quantum resistance. Following Ref. \onlinecite{Bachtold}, we model the NbSe$_3$ nanowires as lossless transmission lines with $Z=\sqrt{l/c}$. The kinetic inductance per unit length $l=\frac{m^*}{e^2 n A}$, where the mass of the charge carriers $m^*=0.24 m_e$ \cite{coleman}, with $m_e$ the electron mass, $e$ the elementary charge. The capacitance per unit length $c=\frac{4\epsilon_0 \epsilon_r}{\pi ln(4d/w)}$, for a rectangular wire above a ground plane at distance $d$, with $\epsilon$ the permittivity ($\epsilon_r=3.9$ for Si-oxide), and $w$ the wire's width. Taking $A=500$~nm$^2$, the kinetic inductance $l=16$~nH/$\mu$m is much larger than the wire's geometrical inductance (pH/$\mu$m). For $d=1~\mu$m and $w=50$~nm, $c \approx 10$~aF/$\mu$m, so that $Z=\sqrt{l/c} \approx 1.6$~R$_Q$ and $\beta=4.2$, in agreement with the exponent measured in the $IV$ at high bias. The assumption of a lossless transmission line is valid when the inductive part $\omega l$ is larger than the resistive part $R/L$ of the impedance, i.e. for high bias: $\hbar \omega = eV > \hbar (R/L)/l \approx 0.4$~meV, for $R/L=10$~k$\Omega /\mu$m. This is consistent with the observed power-law in the $IV$ at high bias.

ECBT may also explain an increase of $\alpha$ for smaller wires, because the impedance depends on the cross section, through the kinetic inductance. Since the capacitance to a ground plane depends only weakly on the wire's width, $\alpha \propto \sqrt{1/A}$ (dashed line in the inset of Fig.~\ref{RT}). Note that this dependency is in agreement with the derived exponent as a function of the number of modes in a quasi-1D wire \cite{glazman}. Although ECBT explains most of our results, its applicability is not straightforward. ECBT is derived for a single tunnel barrier connected to the impedance of the environment, which may not be the case for NbSe$_3$ nanowires. Also, ECBT does not explain the observation $\beta \neq \alpha+1$ and the knee at $eV \approx 80~k_BT$.

An alternative model to describe NbSe$_3$ nanowires is the case of a 1D disordered conductor with low electron density, where Wigner crystallization may occur. Wigner crystallization occurs when the Coulomb energy $E_C$ is larger than the kinetic energy $E_F$ of the electrons or, or in other words, when the coupling parameter $\Gamma_C = E_C/E_F$ is large. In NbSe$_3$, $\Gamma_C$ is large below $T_{P2}$, because the electron density and hence $E_F$ is exceptionally small. A WC can adjusts its phase in the presence of disorder to optimize the pinning energy gain, much like CDW systems or vortex lattices. When the localization length is larger than the distance between impurities, the TDOS follows a power-law \cite{lee}. The power-law exponent is determined by the localization length and high values of 3-6 are predicted, similar to values we found.

In conclusion, we have studied the transport properties of weakly coupled metallic wires in the presence of disorder. We find power-law behavior characteristic for 1D systems when the number of chains is smaller than about 2000. A model to describe NbSe$_3$ nanowires is not available. Ingredients for such a model should include interaction between metallic chains, disorder, confinement and the low electron concentration.

\acknowledgments{We appreciate useful discussions with P.\ H.\ Kes, M.\ Grifoni, Yu.~V.\ Nazarov, and S.~N.~Artemenko. We thank R.~E.\ Thorne for providing the NbSe$_3$ crystals. This work was supported by the Dutch Foundation for Fundamental research on Matter (FOM), the Netherlands Organisation for Scientific Research (NWO), RFBR (project 04-02-16509), and INTAS (project 01-0474). Nanofabrication work was performed at DIMES in Delft.}

\end{document}